\documentclass[aps,prl,twocolumn,a4paper]{revtex4-1}
\usepackage{amsmath}
\usepackage{marvosym}
\usepackage[breaklinks]{hyperref}
\usepackage[utf8]{inputenc}

\newcommand{\exv}[1]{{\langle{#1}\rangle}}
\newcommand{\ket}[1]{{\lvert{#1}\rangle}}
\newcommand{\msym}[1]{{$^\text{#1}$}}

\begin{document}
\noindent
\textbf{Comment on ``State-Independent Experimental Test of Quantum 
Contextuality in an Indivisible System''}

In this Comment we argue that the experiment described in the recent Letter 
 \cite{Zu:2012PRL} does not allow to make conclusions about contextuality.
Our main criticism is that the measurement of the observables as well as the 
 preparation of the state manifestly depend on the chosen context.
Contrary to that, contextuality is about the behavior of the \emph{same} 
 measurement device in different experimental contexts (cf.\ e.g.\ 
 Ref.~\cite{Bell:1966RMP, Peres:1991JPA, Mermin:1993RMP}).

The authors aim to experimentally demonstrate that the noncontextuality 
 assumption is violated by quantum systems.
Specifically, they report a violation of the noncontextuality inequality 
 recently introduced by Yu an Oh \cite{Yu:2012PRL}, which is of the form
\begin{equation}\label{ineq}
 \sum_k \exv{A_k} - \frac14 \sum_{(k,\ell)\in E} \exv{A_k A_{\ell}} \le 8.
\end{equation}
The notation $\exv{A_k A_\ell}$ is an abbreviation denoting the expectation 
 value of the product of the outcomes of the observables $A_k$ and $A_\ell$.
This inequality holds for any noncontextual model, i.e., any model having 
 preassigned values for each observable $A_k$, irrespective of the measurement 
 context (the different pairs $A_kA_\ell$).
Therefore, the experimenter must convincingly argue that the assignment of the 
 observables is independent of the context.
This is a central point in any experimental test of contextuality.
For the argument leading to Eq.~\eqref{ineq} it is thus crucial that (i) the 
 same symbol $A_k$ always corresponds to the same measurement and (ii) the 
 expectation value is evaluated always for the same state of the system.

In Table~\ref{tab}, we list the different measurement procedures that have been 
 used in the experiment, as provided by the supplementary material of the 
 Letter.
Clearly, except for $A_{z_1}$ and $A_{y_3^-}$, none of the observables is 
 measured context independently.
In particular, the observables $A_{h_\alpha}$ ($\alpha=0,1,2,3$) are measured 
 in each context differently, violating condition (i).
In addition the input states are chosen differently for different contexts---an 
 approach that has not been investigated before and directly violates condition 
 (ii).

Since no experimental data or discussion concerning these issues is provided in 
 the Letter, the only means to conclude that those different procedures 
 actually correspond to the same physical observable is to invoke previous 
 knowledge about the functioning of the optical devices.
However, since the setup is operated on a single photon level, this actually 
 requires to employ their quantum mechanical description.
But then the experiment can merely be used to verify the predictions of quantum 
 mechanics \emph{within} the framework of quantum mechanics, rather than a to 
 provide a proof of contextual behavior.

\begin{table}[ttt]
\begin{tabular}{c|c|c|c|c|c|c|c|c|c|c|c|c|c}&
$z_1  $&$z_2  $&$z_3  $&$y_1^-$&$y_2^-$&$y_3^-$&$y_1^+$&
$y_2^+$&$y_3^+$&$h_1  $&$h_2  $&$h_3  $&$h_0  $\\\hline
$z_1  $&1&1&1&1&&&1&&&&&&\\\hline
$z_2  $&1a&?&1a&&3&&&3&&&&&\\\hline
$z_3  $&1a'&1a'&?&&&2a'&&&2a'&&&&\\\hline
$y_1^-$&1b'&&&1b'&&&1b'&&&X2&&&X2\\\hline
$y_2^-$&&3b'&&&3b'&&&3b'&&&Y2&&Y2\\\hline
$y_3^-$&&&2&&&2&&&2&&&2&2\\\hline
$y_1^+$&1b&&&1b&&&1b&&&&X5&X4&\\\hline
$y_2^+$&&3b&&&3b&&&3b&&Y4&&Y5&\\\hline
$y_3^+$&&&2a&&&2a&&&?&4&5&&\\\hline
$h_1  $&&&&X2d&&&&Y4c&4c&4c&&&\\\hline
$h_2  $&&&&&Y2d&&X5c&&5c&&5c&&\\\hline
$h_3  $&&&&&&2d&X4c&Y5c&&&&2d&\\\hline
$h_0  $&&&&X2c&Y2c&2c&&&&&&&2c
\end{tabular}
\caption{\label{tab}%
Different realizations of the 13 observables in the different contexts.
In each row $k$, the entries correspond to the different experimental 
 realizations of the observable $A_k$ depending on the context, i.e., for 
 column $\ell$ in the context $\exv{A_kA_\ell}$, for $\ell=k$ in the single 
 observable context $\exv{A_k}$.
In the entries, the number corresponds to the setting of HWP5 (1: $0^\circ$, 2: 
 $25.5^\circ$, 3: $45^\circ$, 4: $-22.5^\circ$, 5: $67.5^\circ$) and the lower 
 case letter to the setting of HWP6 (a: $0^\circ$, b: $22.5^\circ$, c: 
 $17.63^\circ$, d: $-17.63^\circ$).
Where only the number occurs, the setting of HWP6 does not influence the 
 observable, since the observable was measured using Detector 1; if Detector 3 
 was used, a prime is added.
An $X$ denotes a change of the input state prior to measurement by swapping 
 $\ket0$ and $\ket2$, while $Y$ denotes a swap of $\ket1$ and $\ket2$.
For $\exv{A_{z_2}}$, $\exv{A_{z_3}}$, and $\exv{A_{y_3^+}}$ it is not clear 
 from the material which setting was used in the experiment.}
\end{table}

\newpage
E.~Amselem\msym{\Mercury},
M.~Bourennane\msym{\Mercury},
C.~Budroni\msym{\Venus},
A.~Cabello\msym{\Earth},
O.~Gühne\msym{\Venus},
M.~Kleinmann\msym{\Venus},
J.-Å.~Larsson\msym{\Mars},
and
M.~Wieśniak\msym{\Jupiter}.
\\
\msym{\Mercury}%
Department of Physics, Stockholm University, S-10691 Stockholm, Sweden;
\msym{\Venus}%
Naturwissenschaftlich-Technische Fakultät, Universität Siegen, 
Walter-Flex-Straße 3, D-57068 Siegen, Germany;
\msym{\Earth}%
Departamento de Física Aplicada II, Universidad de Sevilla, E-41012 Sevilla, 
Spain;
\msym{\Mars}%
Institutionen för Systemteknik, Linköpings Universitet, SE-58183 Linköping, 
Sweden;
\msym{\Jupiter}%
Institute of Theoretical Physics and Astrophysics, University of Gdańsk, 80-952 
Gdańsk, Poland.
\bibliography{the}
\end{document}